\documentstyle[epsfig,12pt]{article}

\begin{document}

\hoffset = -1truecm \voffset = -2truecm \baselineskip = 10 mm

\title{\bf From EMC- and Cronin-effects to signals of quark-gluon plasma}

\author{
{\bf Wei Zhu}$^1$, {\bf Jianhong Ruan}$^1$ and {\bf  Fengyao Hou}$^2$\\
\normalsize $^1$Department of Physics, East China Normal University,
Shanghai 200062, P.R. China \\
\normalsize $^2$Kavli Institute for Theoretical Physics China(KITPC),\\
\normalsize Institute of Theoretical Physics, CAS, Beijing 100190, P.R. China}

\date{}

\newpage

\maketitle

\vskip 3truecm

\begin{abstract}

    The EMC- and Cronin-effects are explained by a
unitarized evolution equation, in which the shadowing and antishadowing
corrections are dynamically produced due to gluon fusions. For this
sake, an alternative form of the GLR-MQ-ZRS equation is derived. The
resulting gluon distributions, integrated and unintegrated, in protons and nuclei are
used for analysis of the contributions of the initial parton
distributions to the nuclear suppression factor in heavy ion
collisions. A simulation of the fractional energy loss
is extracted from the data of RHIC and LHC, where the
contributions of both nuclear shadowing and nuclear antishadowing effects are
considered. We find a rapid crossover from weak energy loss to strong energy loss with the gluon jet at a universal critical energy, $E_c\sim 10 GeV$.

\end{abstract}

PACS number(s): 24.85.+p; 12.38.-t; 13.60.Hb

$keywords$: Quark gluon plasma; Nuclear gluon distribution; Energy
loss

\newpage
\begin{center}
\section{Introduction}
\end{center}

    One of the important findings at RHIC and LHC is that the hadron production at high transverse momentum $k_t$
in central heavy ion collisions is suppressed when compared to the
one in p+p collisions [1, 2]. This suppression can be attributed to
energy loss of high-$k_t$ partons that transverse the hot and dense
medium (i.e., quark-gluon plasma QGP) produced in those collisions.
An important goal of the study of heavy ion collisions is therefore
to determine the properties of QGP by measuring the fractional
energy loss where the nuclear effects on the initial parton
distributions should be subtracted.

    The parton densities in a bound nucleon differ from those in a free one.
One example is that the ratio of nuclear structure functions to
deuterium's is smaller or larger than unity at Bjorken variable $x <
0.1$ or $0.1< x < 0.3$. These two facts are called as the nuclear
shadowing and antishadowing in the EMC effect [3]. The nuclear
shadowing and antishadowing effects originate from the gluon fusion
and recombination between two different nucleons in a nucleus, which
will change the distributions of gluon and quarks but not their
total momentum [4]. In consequence, the loss of gluon momentum in
the shadowing range should be compensated by the momentum of new
gluons at larger $x$, which is named the antishadowing effect.

    Another example is the Cronin effect:
the ratio of particle yields in $d+A$ (scaled by the number of
collisions) to those in $p+p$, is over or under unity in an
intermediate transverse momentum range (Cronin enhancement)  or in a
smaller transverse momentum range (anti-Cronin suppression). This
effect was first found at lower fixed target energies [5] and then
was confirmed in $d+Au$ collisions at the BNL Relativistic Heavy Ion
Collider (RHIC)(where $\sqrt{s}= 200 GeV$ )[6-9].

    The Cronin effect is more complicated than the EMC effect. The
former mixes the shadowing-antishadowing corrections at initial
state and the medium modifications at final state. The later provides important information for understanding the properties of dense and hot matter formed in high-energy heavy-ion collisions. Therefore,
the nuclear shadowing and antishadowing effects, which appear
in the EMC effect, should be extracted from the Cronin effect to exposes the properties of the medium.

   The saturation models are broadly used to study the Cronin effect. The saturation is a limiting behavior of the shadowed gluon distribution in the
Jalilian-Marian-Iancu-McLerran-Weigert-Leonidov-Kovner (JIMWLK)
equation [10-15], where the unintegrated gluon distribution is flat
in $k_t$-space when $k_t$ is smaller than the saturation scale $Q_s$
. An elementary QCD process, which also causes nonlinear corrections
in the JIMWLK equation, is the gluon fusion $gg\rightarrow g$. As we
have mentioned before, the antishadowing effect always coexists with
the shadowing effect in \emph{any~gluon~fusion~processes} due to a
general restriction of momentum conservation [16-18]. However, such
antishadowing effect is completely neglected in the original
saturation models. The Cronin enhancement in these models, (i) is
additionally explained as multiple scattering [19-22] using the
Glauber-Mueller model [23] or the McLerran-Venugopalan model
[24,25]; (ii) is produced by special initial gluon distributions of
proton and nucleus [26]. A question is then followed: How much does
the nuclear antishadowing effect contribute to the Cronin effect?

    A global Dokshitzer-Gribov-Lipatov-Altarelli-Parisi (DGLAP) analysis of nuclear parton
distribution functions (for example, the ESP09-set [27,28]) was
proposed. In the DGLAP analysis, the data of Drell-Yan dilepton production from deep inelastic scattering (DIS),
 and the data of inclusive high-$k_t$ hadron
production measured at RHIC are used. It is found that a strong
gluon antishadowing effect is necessary to support the data at RHIC.
Since the DGLAP equation [29-31] does not contain the nonlinear
corrections of gluon fusion, the shadowing and antishadowing effects in the DGLAP
analysis are phenomenologically adopted as initial conditions. However, this brings about uncertainty
due to the lack of the experimental data about nuclear gluon distribution. A similar global DGLAP analysis shows
that the available data are not enough to fix all the complicated input distributions and
that it isn't sure whether the antishadowing
effect does appear or not[32,33]. Besides, the DGLAP equation in the
collinear factorization scheme evolves the integrated parton
distributions. The behavior of the unintegrated gluon distributions,
which contain information of the transverse momentum distribution, is
completely unknown in the DGLAP scheme. Therefore, the ESP09-set of
nuclear parton distributions can't predict with good accuracy the data at lower
$k_t$ at RHIC, where the contributions from intrinsic transverse momentum
become more important [27,28].

   The modification of the gluon recombination to the standard DGLAP
evolution equation was first proposed by Gribov-Levin-Ryskin and
Mueller-Qiu (the GLR-MQ equation) in [34,35]. This GLR-MQ equation is naturally
regarded as a better scheme to describe the QCD dynamics of the nuclear shadowing since the same
gluon fusion exists both in proton and in nucleus but differs in the strength of the nonlinear terms [36,37].
However, the
GLR-MQ equation can't predict the nuclear antishadowing effect
due to it violates the momentum conservation. This
defect is remedied by a modified equation
(the GLR-MQ-ZRS equation) proposed by Zhu, Ruan and Shen in their works[38-40], where
the corrections of the gluon fusion to the DGLAP equation lead to
both shadowing and antishadowing effects. Although, the integral
solutions of the GLR-like equations in present need the initial
distributions on a boundary line $(x,Q^2_0)$ at a fixed $Q^2_0$, and
the unknown input with nuclear shadowing and antishadowing effects at small $x$ still exist.

    This work tries to improve the GLR-like methods motioned above. We study the nuclear shadowing and antishadowing effects in the EMC- and Cronin-effects, which are dynamically arisen from the gluon recombination. Then we use the resulting nuclear gluon distributions to produce the contributions of the initial parton
distributions to the nuclear suppression factor in heavy ion
collisions and to extract fractional energy loss from the data in RHIC and in LHC.
For this sake, an alternative form of the GLR-MQ-ZRS equation at the
double-leading-logarithmic-approximation (DLLA) is derived in Sec. 2. This equation will evolve along small $x$-direction.
The nonlinear corrections to the input distributions can be neglected
if the value of the starting point $x_0$ is large enough. Both the
shadowing and antishadowing effects are naturally grown up with the
evolution of $x$ along the direction from $x_0$ to smaller $x$. This scheme avoids the non-perturbative nuclear
modifications to the input distributions and then simplifies the initial
conditions. The existing data about the EMC- and Cronin-effects is used to fix a few of free parameters in the solutions. Then the integrated and unintegrated gluon distributions in proton and nuclei are obtained to analyze the nuclear suppression factor in heavy ion collisions.

    Our main conclusions are: (i) we support the stronger
shadowing-antishadowing effects both in the unintegrated and
integrated gluon distributions due to a strong $A$-dependence of the
nonlinear corrections in the heavy nucleus; (ii) both the
anti-Cronin suppression and Cronin enhancement mainly originate from
the same gluon recombination mechanism in the nuclear shadowing and
antishadowing effects of the EMC effect; (iii) fractional energy
loss is rapidly crossover from weak energy loss to strong energy
loss with the gluon jet at a universal critical energy, $E_c\sim 10
GeV$.

    This work is organized as follows. We derive
the GLR-MQ-ZRS equation in a new form in Sec.2. Basing on this equation
we study the shadowing and antishadowing effects in the EMC effect
in Sec.3. The nuclear shadowing and antishadowing contributions to the Cronin effect are exposed by using the resulting unintegrated gluon distributions in proton and nuclei in Sec.4. The
nuclear shadowing and antishadowing effects to the nuclear
suppression factor are predicted and a simulations of
fractional energy loss is extracted from the data at RHIC and at LHC in Sec. 5.

\newpage
\begin{center}
\section{A new form of the GLR-MQ-ZRS equation}
\end{center}

    In history, the DGLAP evolution equation is derived by using the renormalization group
method for the integrated distributions. The resulting equation
evolves with factorization scale $\mu$. In this section, we try to
rewrite the DGLAP equation with the nonlinear modifications
beginning from the unintegrated distributions. Then, we get an
alternative form of the equation, which evolves the Bjorken variable
$x$.

    We begin from a deep inelastic scattering process, where
the unintegrated gluon distribution is measured. In the
$k_t$-factorization scheme, the cross section is decomposed into

$$d\sigma(probe^* P\rightarrow kX)$$
$$=f(x_1,k_{1t}^2)\otimes{\cal
K}\left(\frac{k_t^2}{k_{1t}^2},\frac{x}{x_1},\alpha_s\right)\otimes
d\sigma(probe^*k_1\rightarrow k)$$
$$\equiv \Delta f(x,k_t^2)\otimes d\sigma(probe^*k_1\rightarrow
k), \eqno(1)$$ which contains the evolution kernel ${\cal K}$, the
unintegrated gluon distribution function $f$ and the
$probe^*$-parton cross section $d\sigma(probe^*k_1\rightarrow k)$.
For simplicity, we fix the QCD coupling at the leading order (LO)
approximation in this work. According to the scale-invariant parton
picture of the renormalization group [41], we regard $\Delta
f(x,k_t^2)$ as the increment of the distribution $f(x_1,k_{1t}^2)$
when it evolves from $(x_1,k_{1t}^2)$ to $(x,k_t^2)$. Thus, the
connection between $f(x_1,k^2_{1t})$ and $f(x,k^2_t)$ via Eq. (1) is

$$f(x,k_t^2)=f(x_1,k_{1t}^2)+\Delta
f(x,k_t^2)$$
$$=f(x_1,k_{1t}^2)+\int^{k_t^2}_{k^2_{1t,{min}}}\frac{dk^2_{1t}}{k^2_{1t}}
\int^1_{x}\frac{dx_1}{x_1} {\cal
K}\left(\frac{k_t^2}{k_{1t}^2},\frac{x}{x_1},\alpha_s\right)
f(x_1,k_{1t}^2). \eqno(2) $$ In the case of the LO DGLAP evolution,
we adopt a physical gauge(axial gauge), in which only the
transverse gluon polarizations are summed over , so that the ladder-type diagrams dominate the evolution. The unintegrated distributions  satisfy the
normalization relation

$$G(x,\mu^2)\equiv
xg(x,\mu^2)=\int^{\mu^2}_{k^2_{t,min}}\frac{d k_t^2}{k^2_t}
xf(x,k_t^2)\equiv\int^{\mu^2}_{k^2_{t,min}}\frac{dk^2_t}{k_t^2}
F(x,k_t^2), \eqno(3)$$ where the possible non-logarithmic tail when
$k_t>\mu$ are beyond NLO accuracy. These distributions correspond to
the density of partons in the proton with longitudinal momentum
fraction $x$ with the parton transverse momentum integrated up to
$k_t=\mu$.

   From Eqs. (2) and (3), we have

$$\Delta g(x,\mu^2)=\int^{\mu^2}_{k^2_{t,min}}
\frac{dk^2_t}{k^2_t}
\int^{k_t^2}_{k^2_{1t,{min}}}\frac{dk^2_{1t}}{k^2_{1t}}
\int_x^1\frac{dx_1}{x_1}\frac{1}{x_1}{\cal K}\left
(\frac{k_t^2}{k_{1t}^2},\frac{x}{x_1},\alpha_s\right)F(x_1,k_{1t}^2),\eqno(4)$$
or

$$\Delta G(x,\mu^2)=\Delta xg(x,\mu^2)=
\int^{\mu^2}_{k^2_{T,min}} \frac{dk^2_t}{k^2_t}
\int^{k_t^2}_{k^2_{1t,{min}}}\frac{dk^2_{1t}}{k^2_{1t}}
\int_x^1\frac{dx_1}{x_1}\frac{x}{x_1}{\cal K}\left
(\frac{k_t^2}{k_{1t}^2},\frac{x}{x_1},\alpha_s\right)F(x_1,k_{1t}^2)$$
$$=\int^{\mu^2}_{k^2_{t,min}}\frac{dk^2_t}{k^2_t}
\int_x^1\frac{dx_1}{x_1}\frac{x}{x_1}{\cal K}\left
(\frac{x}{x_1},\alpha_s\right)G(x_1,k^2_{t}),\eqno(5)$$ where the
last step is valid when the $k_t$ is strongly ordered. Usually DGLAP
evolution equation is written with the virtuality $\mu^2$ rather
than with $k^2_t$ , but at LO level the equation is the same with
these two different arguments since the difference between them is a
NLO effect. Therefore, we take

$$G(x,k^2_t)\rightarrow G(x,\mu^2).\eqno(6)$$ Thus, in

$$G(x,\mu^2)=G(x_1,\mu^2_1)+\Delta G(x,\mu^2),\eqno(7)$$ we write

$$\Delta G(x,\mu^2)$$
$$=\int^{\mu^2}_{\mu_{1,min}^2}\frac{d\mu_1^2}{\mu_1^2}\int^1_x\frac{dx_1}{x_1}\frac{x}{x_1}
{\cal
K}_{DGLAP}^{LL(\mu^2)}(\frac{x}{x_1},\alpha_s)G(x_1,\mu^2_1),\eqno(8)$$
at the leading logarithmic $\mu^2$ approximation, in which the
unregularized splitting kernel

$$\frac{dx_1}{x_1}{\cal K}_{DGLAP}^{LL(\mu^2)}
=\frac{\alpha_sN_c}{\pi}
\frac{dx_1}{x_1}[z(1-z)+\frac{1-z}{z}+\frac{z}{1-z}]$$
$$=\frac{\alpha_sN_c}{\pi}
\frac{dx_1}{x_1}\left[\frac{x(x_1-x)}{x_1^2}+\frac{x_1-x}{x}+\frac{x}{x_1-x}\right]\eqno(9a)$$
$$\stackrel{x\ll x_1}{\longrightarrow}\frac{dx_1}{x_1}
{\cal K}_{DGLAP}^{DLL} =\frac{\alpha_sN_c}{\pi} \frac{dx_1}{x}.
\eqno(9b)$$

     We add the contributions of the nonlinear recombination
terms in Eq. (8) according to Refs. [38-40] (See Appendix),

$$\Delta G(x,\mu^2)$$
$$=\int^{\mu^2}_{\mu_{1,min}^2}\frac{d\mu_1^2}{\mu_1^2}\int^1_x\frac{dx_1}{x_1}\frac{x}{x_1}
{\cal K}_{DGLAP}^{LL(\mu^2)}(\frac{x}{x_1},\alpha_s)G(x_1,\mu^2_1)$$
$$-2\int^{Q^2}_{\mu^2_{1min}}\frac{d\mu^2_1}{\mu^4_1}\int_{x}^{1/2}\frac{dx_1}{x_1}\frac{x}{x_1}
{\cal K}_{GLR-MQ-ZRS}^{GG\rightarrow
GG,~LL(\mu^2)}\left(\frac{x}{x_1},\alpha_s\right)
G^{(2)}(x_1,\mu_1^2)$$
$$+\int^{\mu^2}_{\mu^2_{1min}}\frac{d\mu^2_1}{\mu^4_1}\int_{x/2}^{1/2}\frac{dx_1}{x_1}\frac{x}{x_1} {\cal
K}_{GLR-MQ-ZRS}^{GG\rightarrow
GG,~LL(\mu^2)}\left(\frac{x}{x_1},\alpha_s\right)
G^{(2)}(x_1,\mu_1^2),\eqno(10)$$ where

$$\frac{dx_1}{x_1}{\cal K}_{GLR-MQ-ZRS}^{GG\rightarrow GG,~LL(\mu^2)}$$
$$=\frac{\alpha^2_s}{8} \frac{N^2_c}{N^2_c-1}\frac{(2x_1-x)(72x_1^4-48x_1^3x+140x_1^2x^2-116x_1x^3+29x^4)}{x_1^5x}dx_1
\eqno(11a)$$
$$\stackrel{x\ll x_1}{\longrightarrow}\frac{dx_1}{x_1}
{\cal K}_{GLR-MQ-ZRS}^{GG\rightarrow GG,~DLL}=
18\alpha^2_s\frac{N^2_c}{N^2_c-1}\frac{dx_1}{x}. \eqno(11b)$$

$$G^{(2)}(x,\mu^2)=R_GG^2(x,\mu^2),\eqno(12)$$ where $R_G=1/(\pi R^2)$ is a correlation coefficient with the dimension
$[L^{-2}]$, and $R$ is the effective correlation length of two
recombination gluons. One can easily get the GLR-MQ-ZRS equation at
DLL approximation

$$\frac{\partial G(x,\mu^2)}{\partial\ln \mu^2}$$
$$=\frac{\alpha_sN_c}{\pi}\int^1_x
\frac{dx_1}{x_1}G(x_1,\mu^2)-\frac{36\alpha_s^2}{\pi
\mu^2R^2}\frac{N^2_c}{N^2_c-1}
\int_x^{1/2}\frac{dx_1}{x_1}G^2(x_1,\mu^2)
$$
$$+\frac{18\alpha_s^2}{\pi
\mu^2R^2}\frac{N_c^2}{N_c^2-1}
\int_{x/2}^{1/2}\frac{dx_1}{x_1}G^2(x_1,\mu^2). \eqno(13)$$

    It is interesting to compare this small-$x$ version of the GLR-MQ-ZRS
equation with the GLR-MQ equation, which is [35]

$$\frac{\partial G(x,\mu^2)}{\partial\ln \mu^2}$$
$$=\frac{\alpha_sN_c}{\pi}\int^1_x
\frac{dx_1}{x_1}G(x_1,\mu^2)-\frac{36\alpha_s^2}{8
\mu^2R^2}\frac{N^2_c}{N^2_c-1}
\int_x^{1/2}\frac{dx_1}{x_1}G^2(x_1,\mu^2),  \eqno(14)$$ where

$$G^{(2)}(x,\mu^2)=\frac{9}{8\pi R^2}G^2(x,\mu^2),\eqno(15)$$ is assumed.

    Comparing with the GLR-MQ equation (14), there are several features in
the GLR-MQ-ZRS equation (13): (i) the momentum conservation of
partons is maintained; (ii) because of the shadowing and
antishadowing effects in Eq. (13) have different kinematic regions,
the net effect depends both on the local value of the gluon
distribution at the observed point and on the shape of the
gluon distribution when the Bjorken variable goes from $x$ to $x/2$.
In consequence, the shadowing effect in the evolution process will
be obviously weakened more by the antishadowing effect if the
distribution is steeper. Therefore, the antishadowing effect can not
be neglected in the pre-saturation range.

    According to the definition Eq. (3), one can roughly estimate the
unintegrated gluon distribution using

$$ F(x,k_t^2)\simeq\left.\mu^2\frac{\partial  G(x,\mu^2)} {\partial \mu^2}
\right\vert_{\mu^2=k_t^2}.\eqno(16)$$ However, Eq. (16) will be
invalid with $x$ increases, since the contribution of negative
virtual DGLAP term will exceed the contribution of real emission one
and lead to negative values of $F$. In fact, due to strong $k_t$
ordered in DGLAP evolution, the transverse momentum of the final
parton to leading-order is obtained just at the last step of the
evolution. Thus, the $ k_t$-dependent distribution can be calculated
directly from the DGLAP equation if only keeping the contribution of
a single real emission. Meanwhile, all the virtual contributions
from the scale of $k_t$ up to the final scale $\mu$ of the hard
subprocess are resummed up into a Sudakov factor T, which describes
the probability of no parton emission during the evolution. However,
at the small $x$ range, the virtual contributions to the gluon
distribution in the DGLAP kernel can be neglected. We have indicated
that the contributions of the virtual processes in the nonlinear
kernels of the GLR-MQ-ZRS equation are canceled [38], therefore, the
Sudakov form factors in nucleon are the same as those in nucleus[42]
and can be canceled in their ratio. Thus, we use

$$\Delta G(x,\mu^2)$$
$$=\int^{\mu^2}_{k_{min}^2}\frac{dk_t^2}{k_t^2}
\int^{k_t^2}_{k^2_{1t,min}} \frac{dk^2_{1t}}{k^2_{1t}}
\int^1_x\frac{dx_1}{x_1}\frac{x}{x_1} {\cal
K}_{DGLAP}^{DLL}(\frac{x}{x_1},\alpha_s)F(x_1,k_{1t}^2)$$
$$-2\int^{\mu^2}_{\mu^2_{1min}}\frac{dk^2_t}{k_t^4}\
\int_{x}^{1/2}\frac{dx_1}{x_1}\frac{x}{x_1} {\cal
K}_{MD-DGLAP}^{DLL}\left(\frac{x}{x_1},\alpha_s\right)
G^{(2)}(x_1,k^2_t)$$
$$+\int^{\mu^2}_{\mu^2_{1min}}\frac{dk_t^2}{k_t^4}
\int_{x/2}^{1/2}\frac{dx_1}{x_1}\frac{x}{x_1} {\cal
K}_{MD-DGLAP}^{DLL}\left(\frac{x}{x_1},\alpha_s\right)
G^{(2)}(x_1,k_t^2),\eqno(17)$$ and obtain

$$\Delta F(x,k_t^2)=\left.\mu^2\frac{\partial\Delta G(x,\mu^2)} {\partial \mu^2}
\right\vert_{\mu^2=k_t^2}$$
$$=\frac{\alpha_sN_c}{\pi}\int^{k^2_t}_{k^2_{1t,min}}\frac{dk^2_{1t}}{k^2_{1t}}\int^1_x\frac{dx_1}{x_1}F(x_1,k_{1t}^2)$$
$$-\frac{36\alpha_s^2}{k^2_t}\frac{N_c^2}{N_c^2-1}
\int_{x} ^{1/2}\frac{dx_1}{x_1}G^{(2)}(x_1,k_t^2)$$
$$+\frac{18\alpha_s^2}{k^2_t}\frac{N_c^2}{N_c^2-1}
\int_{x/2} ^{1/2}\frac{dx_1}{x_1}G^{(2)}(x_1,k_t^2), \eqno(18)$$

    Now we re-derive the GLR-MQ-ZRS equation, which evolves with the longitudinal momentum now. We differentiate

$$F(x,k_t^2)=F(x_1,k_{1t}^2)+\Delta F(x,k_t^2),\eqno(19)$$ with respect to
$x$. Note that

$$-\frac{\partial F(x,k_t^2)}{\partial x}$$
$$=\left.\int^{k_t^2}_{k^2_{1t,min}}
\frac{dk^2_{1t}}{k_{1t}^2}\frac{1}{x_1} {\cal
K}\left(\frac{k_t^2}{k_{1t}^2},\frac{x}{x_1},\alpha_s\right)
F(x_1,k_{1t}^2) \right | _{x_1=x}$$
$$-\int^{k_t^2}_{k^2_{1t,mi}}
\frac{dk_{1t}^2}{k_{1t}^2}
\int^1_{x}\frac{dx_1}{x_1^2}\frac{\partial {x\cal
K}\left(\frac{k_t^2}{k_{1t}^2},\frac{x}{x_1},\alpha_s\right)}{\partial
x} F(x_1,k_{1t}^2). \eqno(20)$$ Generally, the resummation solution
is hard to be obtained from this equation. However, the second term
on the right-hand side of Eq. (20) vanishes at the $LL(1/x)$
approximation . In this case, the resummation becomes simple, i.e.,
we have

$$-x\frac{\partial F(x,k_t^2)}{\partial x}$$
$$=\frac{\alpha_{s}N_c}{\pi}\int^{k_t^2}_{k_{1t,min}}\frac{d
k_{1t}^2}{k_{1t}^2}F(x_,k_{1t}^2)$$
$$-\frac{36\alpha_s^2}{\pi
R^2k_t^2}\frac{N_c^2}{N_c^2-1}\left[\int^{k_t^2}_{k_{1t,min}^2}\frac{d
k_{1t}^2}{k_{1t}^2} F(x,k_{1t}^2)\right]^2+\frac{18\alpha_s^2}{\pi
R^2k_t^2}\frac{N_c^2}{N_c^2-1}\left[\int^{k_t^2}_{k_{1t,min}^2}\frac{d
k_{1t}^2}{k_{1t}^2} F\left(\frac{x}{2},k_{1t}^2\right)\right]^2,$$
$$(x\le 0.15);
$$

$$-x\frac{\partial F(x,k_t^2)}{\partial x}$$
$$=\frac{\alpha_{s}N_c}{\pi}\int^{k_t^2}_{k_{1t,min}}\frac{d
k_{1t}^2}{k_{1t}^2}F(x_,k_{1t}^2)+\frac{18\alpha_s^2}{\pi
R^2k_t^2}\frac{N_c^2}{N_c^2-1}\left[\int^{k_t^2}_{k_{1t,min}^2}\frac{d
k_{1t}^2}{k_{1t}^2}
F\left(\frac{x}{2},k_{1t}^2\right)\right]^2,~~(0.15\le x\le 0.3),
\eqno(21)$$ This is a new form of the GRL-MQ-ZRS equation. The
negative and positive nonlinear terms correspond to the shadowing
and antishadowing effects in the gluon recombination. Here the
shadowing and antishadowing coexist in the region $x\le 0.15$, while
only the antishadowing  exits in $0.15\le x\le 0.3$.

 The kinematic regions of Eq. (21) can be explained as follow.
The evolution kernel of the GLR-MQ-ZRS equation (11a) as same as the DGLAP equation (9a) is derived at
LL$(Q^2)$ approximation and valid in a whole $x$ range. However the
DLLA form of the GLR-MQ-ZRS equation (11b) is valid at the small $x$
range when $x<x_1$, here $x_1 \sim {\cal {O}}(10^{-1})$ according
to $\alpha_s\ln(1/x)\ln (k^2_T/\mu^2)\sim{\cal {O}}(1)$. We take
$x_1=0.15$ in this work.

     Now we apply Eq. (21) in the nuclear target.
The Shadowing and antishadowing effects thought arise from a
nonlinear mechanism when gluons are sufficiently dense
to interact with themselves. The strength of the gluon recombination is
proportional to the gluon density in the transverse area. The gluons
with smaller $x$ will exceed the longitudinal size of nucleon in a
nucleus. Thus, the strength of the nonlinear recombination terms in
Eq. (21) should be scaled by $A^{1/3}$ in a nucleus. On the other
hand, although the softer gluons of different nucleons with extra
small $k_t$ maybe correlate with each other in the transverse area,
we still neglect these corrections because the
integrations on $k_t$ can go down to a very small value in Eq. (21) and $F(x,k_t^2\rightarrow
0)\rightarrow 0$ . In this simple model, Eq. (21) in the nucleus
becomes

$$-x\frac{\partial F_A(x,k_t^2)}{\partial x}$$
$$=\frac{\alpha_{s}N_c}{\pi}\int^{k_t^2}_{k_{1t,min}^2}\frac{d
k_{1t}^2}{k_{1t}^2}F_A(x_,k_{1t}^2)$$
$$-A^{1/3}\frac{36\alpha_s^2}{\pi
R^2k_t^2}\frac{N_c^2}{N_c^2-1}\left[\int^{k_t^2}_{k_{1t,min}^2}\frac{d
k_{1t}^2}{k_{1t}^2}
F_A(x,k_{1t}^2)\right]^2+A^{1/3}\frac{18\alpha_s^2}{\pi
R^2k_t^2}\frac{N_c^2}{N_c^2-1}\left[\int^{k_t^2}_{k_{1t,min}^2}\frac{d
k_{1t}^2}{k_{1t}^2} F_A\left(\frac{x}{2},k_{1t}^2\right)\right]^2,$$
$$(10^{-2}\le x\le 0.15); $$

$$-x\frac{\partial F_A(x,k_t^2)}{\partial x}$$
$$=\frac{\alpha_{s}N_c}{\pi}\int^{k_t^2}_{k_{1t,min}^2}\frac{d
k_{1t}^2}{k_{1t}^2}F_A(x_,k_{1t}^2)+A^{1/3}\frac{18\alpha_s^2}{\pi
R^2k_t^2}\frac{N_c^2}{N_c^2-1}\left[\int^{k_t^2}_{k_{1t,min}^2}\frac{d
k_{1t}^2}{k_{1t}^2}
F_A\left(\frac{x}{2},k_{1t}^2\right)\right]^2,~~(0.15\le x\le 0.3),
\eqno(22)$$

    The input gluons are distributed on the boundary line
$(2x_0,k_t)$ at fixed $2x_0$. We take a larger value of $x_0$ as a
starting point of the evolution, where the gluons just begin to fuse
and the nonlinear corrections to the input gluon distributions can
be neglected. Meanwhile the contributions of the Fermi
motion and nuclear binding effects to the nuclear parton
distributions at small $x$ are small enough. In this case, a nucleus is
composed simply with incoherent constituent nucleons at $2x_0$. we have

$$
F(2x_0,k^2_t)=F_A(2x_0,k^2_t),\eqno(23)$$ where the nuclear parton
distributions have been normalized.

   Although a pure DGLAP equation is used at very small $x$ in some papers,
we find the DGLAP equation with shadowing corrections Eq. (21)
predicts a smaller $F_2$ than experiment data when $x<10^{-3}$. In
fact, the DGLAP kernel in Eq. (21) resums the leading $\alpha_s
\ln(1/x)\ln(k^2_t/\mu ^2)$ contributions doubly. As we know that the
BFKL evolution [43-48] which resums the leading $\ln(1/x)$
contributions and should replace the DGLAP equation
at very small $x$ ($x<10^{-3}$) according to
$\alpha_s\ln (1/x)\sim{ \cal {O}}(1)$. In this region, we write Eq.
(2) as

$$\Delta
f(x,k_t^2)=\int^{\infty}_{k^2_{1t,{min}}}\frac{d^2{\bf{k}}_{1t}}{k^2_{1t}}
\int^1_{x}\frac{dx_1}{x_1} {\cal
K}\left(\frac{k_t^2}{k_{1t}^2},\frac{x}{x_1},\alpha_s\right)
f(x_1,k_{1t}^2), \eqno(24) $$ or

$$\Delta
F(x,k_t^2)=\int^{\infty}_{k^2_{1t,{min}}}\frac{d^2{\bf{k}}_{1t}}{k^2_{1t}}
\int^1_{x}\frac{dx_1}{x_1}\frac{x}{x_1} {\cal
K}\left(\frac{k_t^2}{k_{1t}^2},\frac{x}{x_1},\alpha_s\right)
F(x_1,k_{1t}^2), \eqno(25) $$ where

$$F(x,k_t^2)=F(x_1,k_{1t}^2)+\Delta
F(x,k_t^2), \eqno(26) $$ and

$$\frac{x}{x_1}{\cal
K}\left(\frac{k_t^2}{k_{1t}^2},\frac{x}{x_1},\alpha_s\right)\rightarrow
{\cal
K}_{BFKL}^{LL(1/x)}\left(\frac{k_t^2}{k_{1t}^2},\alpha_s\right).
\eqno(27)$$ In consequence, we have the linear BFKL equation.

$$-x\frac{\partial F(x,k_t^2)}{\partial x}$$
$$=\frac{\alpha_{s}N_ck_t^2}{\pi}\int_{k^2_{1t,min}}^{\infty} \frac{d {k}_{1t}^2}{k^2_{1t}}
\left\{\frac{F(x,k_{1t}^2)-F(x,k_t^2)} {\vert
k^2_{1t}-k^2_t\vert}+\frac{F(x,k_t^2)}{\sqrt{k^4_t+4k^4_{1t}}}\right\}
,\eqno(28)$$ and nonlinear BFKL equation with the modifications of
gluon fusion

$$-x\frac{\partial F(x,k_t^2)}{\partial x}$$
$$=\frac{\alpha_{s}N_ck_t^2}{\pi}\int_{k^2_{1t,min}}^{\infty} \frac{d {k}_{1t}^2}{k^2_{1t}}
\left\{\frac{F(x,k_{1t}^2)-F(x,k_t^2)} {\vert
k^2_{1t}-k^2_t\vert}+\frac{F(x,k_t^2)}{\sqrt{k^4_t+4k^4_{1t}}}\right\}$$
$$-\frac{36\alpha_s^2}{\pi
R^2k_t^2}\frac{N_c^2}{N_c^2-1}\left[\int^{k_t^2}_{k_{1t,min}^2}\frac{d
k^2_{1t}}{k_{1t}^2} F(x,k_{1t}^2)\right]^2+\frac{18\alpha_s^2}{\pi
R^2k_t^2}\frac{N_c^2}{N_c^2-1}\left[\int^{k_t^2}_{k_{1t,min}^2}\frac{d
k^2_{1t}}{k_{1t}^2} F\left(\frac{x}{2},k_{1t}^2\right)\right]^2;
$$

$$-x\frac{\partial F_A(x,k_t^2)}{\partial x}$$
$$=\frac{\alpha_{s}N_ck_t^2}{\pi}\int_{k^2_{1t,min}}^{\infty} \frac{d {k}_{1t}^2}{k^2_{1t}}
\left\{\frac{F_A(x,k_{1t}^2)-F_A(x,k_t^2)} {\vert
k^2_{1t}-k^2_t\vert}+\frac{F_A(x,k_t^2)}{\sqrt{k^4_t+4k^4_{1t}}}\right\}$$
$$-A^{1/3}\frac{36\alpha_s^2}{\pi
R^2k_t^2}\frac{N_c^2}{N_c^2-1}\left[\int^{k_t^2}_{k_{1t,min}^2}\frac{d
k^2_{1t}}{k_{1t}^2}
F_A(x,k_{1t}^2)\right]^2+A^{1/3}\frac{18\alpha_s^2}{\pi
R^2k_t^2}\frac{N_c^2}{N_c^2-1}\left[\int^{k_t^2}_{k_{1t,min}^2}\frac{d
k^2_{1t}}{k_{1t}^2}
F_A\left(\frac{x}{2},k_{1t}^2\right)\right]^2.\eqno(29)$$

    To solve the equations numerically we need to know the mix of the BFKL- and DGLAP-equatons. A
unified framework which works in all the through $(x,k_t)$ kinematic
region was provided by Catani, Ciafaloni, Fiorani and Marchesini
(the CCFM equation [49-51]). Based on the coherent radiation of
gluons, this equation leads to an angular ordering of the gluon
emissions along the chain. In the leading order approximation of
$\ln(1/x)$, the CCFM equation reduces to the BFKL equation, whereas
at moderate $x$ the angular ordering becomes an ordering in gluon
transverse momenta and the CCFM equation becomes equivalent to the
standard DGLAP equation.

    Unfortunately, in the CCFM schema  there contains unknown shadowing and
antishadowing information in its complicate input distributions. As
we know that the BFKL equation can be derived in a dipole picture
[52-56]. At large $x$ region ($x>0.1$), parton densities in nucleon
are dilute and the probe interacts with a single parton (Fig.1 a).
In this case, the DGLAP dynamics are dominant. At smaller $x$ region
($x<10^{-3})$, the correlations among the initial partons in a
nucleon becomes more important and the dipole configuration
dominates the initial state (Fig.1c), and the BFKL dynamics in place
of the DGLAP dynamics are dominant. Note that although when
$x<10^{-3}$ the BFKL evolution is dominant according to
$\alpha_s\ln(1/x)\sim {\cal {O}}(1)$, we can not exclude directly
the BFKL dynamics from the $part$ of evolution at a larger $x$
region according to Fig. 1b. A natural connection between two
evolution dynamics should be that the BFKL dynamics replace
asymptotically the DGLAP dynamics from $x=0.1$ to $10^{-3}$ through
out the mixing region of the single parton and dipole configuration.
Concretely, we take

$$Equation (21), ~~at~0.15>x>10^{-1};  $$
$$Equation (21) \times [1-\beta]+Equatiuion (29) \times \beta, ~at~10^{-1}>x>10^{-3};$$
$$Equation (29),~~at~x<10^{-3},\eqno(30)$$ where

$$\beta=-\frac{\ln 10}{\ln 10^3-\ln 10}+\frac{\ln\frac{1}{x}}{\ln 10^3-\ln 10}.\eqno(31)$$

    We emphasize that if we use the original form of the GLR-MQ-ZRS
equation (13) to replace Eq. (21), the solution of the Eq. (31)
becomes very difficult to solve since there exists two different evolution ways.

     All the parameters in the solutions of Eqs. (21), (22) and (29) will be determined by the EMC effect.
Most of the data about the EMC effect are got by measuring the structure
functions. Thus, we should calculate the quark distributions at
small $x$. We assume that the sea quark distributions at the small
$x$ range are dominant compared with the gluon distribution, via the DGLAP
splitting process $g\rightarrow q\overline{q}$. Thus, the structure function of the deep inelastic procress at small $x$ reads [57]

    $$F_2(x,Q^2)$$
$$=2\int^1_xdx_1\int^{Q^2}\frac{dk^2_t}{k^2_t}
\int^{k^2_t}\frac{dk^2_{1t}}{k^2_{1t}}F(\frac{x}{x_1},k^2_{1t})
\sum_qe^2_q\frac{\alpha_s}{2\pi}P_{qg}(x_1). \eqno(32)$$ where
$P_{qg}(x_1)$ is the DGLAP splitting function.

\newpage
\begin{center}
\section{The EMC effect}
\end{center}

    The EMC effect includes the Fermi motion and
binding effect at $x>0.3$ [3]. However,
in this work we focus the nuclear shadowing and antishadowing contributions at $x<0.3$ since we use the
RHIC and LHC data at $x\sim k_t/\sqrt{s}<0.3$.

       We choose $x_0=0.15$ as the starting
point of the evolution in Eqs. (21) and (22), where the nonlinear
gluon recombination begins to work. We find that following input is
suitable, i.e.,

$$F(2x_0,k_t^2)=2\sqrt{(k_t^2)}exp(-(log(k_t^2))^2). \eqno(33)$$

   It is necessary to know the value of $F(x_i/2,k^2_t)$ at the step $x=x_i$ in advance to compute Eqs. (21), (22) and (29) numerically. For this sake, we proposed the following program in [58]

$$F\left(\frac{x_i}{2},k^2_t\right)=F_{Shadowing}\left(\frac{x_i}{2},k^2_t\right)
+\frac{F_{linear}\left(\frac{x_i}{2},k^2_t\right)-F_{Shadowing}\left(\frac{x_i}{2},k^2_t\right)}{i\Delta
-\Delta+1} ,\eqno(34)$$ where $F_{Shadowing}(x_i/2,k^2_t)$ (or
$F_{linear}(x_i/2,k^2_t)$) indicates that the evolution from $x_i$
to $x_i/2$ is controlled by Eqs. (21), (22) and (29) without the
antishadowing contributions (or is controlled by the linear
equation). The parameter $0<\Delta<1$, which implies the different
velocities approaching to the dynamics of Eqs. (21), (22) and (29).

       At first, we use the well known data of $F_{2p}(x,Q^2)$ [59,60] of a free proton in order to
determine the parameters in the computations. Then we predict the
distributions in nuclei. The dashed curve in Fig. 2 is our fitting
result using the input (33), where we take the parameters $R=2.4$
$GeV^{-1}$, $k_t^2=0.01 GeV^2$, $\alpha_s=0.3$ and $\Delta=0.02$.
Note that the contributions of the valence quarks to $F_2$ at
$x>0.1$ are necessary and they can be parameterized by the
difference between the dashed curve and experimental solid curve
in Fig.2.

    Figure 3 shows our predictions of Eqs. (21), (22) and (29) for the
Ca/C, C/Li, Ca/D and Cu/D compared with the EMC and NMC results
[61-63]. Their agreement is acceptable.

    Different from the scheme of the DGLAP evolution equation, our scheme can predict
the nuclear effects for the unintegrated gluon distribution. The
results are presented by their ratios of the nuclear unintegrated
gluon distributions in Figs.4 and 5. Although deep inelastic
scattering experiments do not examine directly these
effects, hadron-nucleus scattering at RHIC relates the nuclear
unintegrated gluon distributions, which will be used in next
section.

   Using Eqs. (21), (22) and (29), the ratios of gluon distributions$G_{Ca}(x,Q^2)/G_D(x,Q^2)$ at
$Q^2=2$ and $10~GeV^2$ are given in Fig.6. The curves present a
cusps at $x=0.15$. This is due to a simply assumption in Eq. (12),
which leads to the shadowing and antishadowing effects start from
$x=0.15$ and $x=0.3$, respectively in Eq. (22). One can smear the
cusps when considering the gluon fusions with different values of
$x$. However, this will complicate the calculations but doesn't
change our following conclusions.

    The $Q^2$-dependence of the gluon ratio is predicted in the region
$10^{-4}<x<10^{-1}$ in our model. The logarithmic slope $b$ in
$G_A/G_{A'}=a+b\ln Q^2$ is positive. However, the corresponding
slope in the ratio of the structure functions $F_{2Ca}/F_{2D}$ is
negative at small $x$ (see Fig.7). For example, the predicted
$Q^2$-slope for calcium at $x\simeq 4\times10^{-2}$, $b\simeq
-0.046$, and at $x\simeq 10^{-2}$, $b\simeq -0.03$, the results are
compatible with the measured data in [64].

    The data of $G_{Sn}(x,Q^2)/G_C(x,Q^2)$ are measured from
inelastic $J/\psi$ production by the NM Collaboration in Ref. [65],
which determine a stronger nuclear antishadowing effect but with a
larger uncertainty. Our prediction is presented in Fig.8.

     Compared with the ESP09 set [27,28], our works predict a more stronger antishadowing effect
in the gluon (integrated and unintegrated) distributions of heavy
nucleus. One reasons is that the observed antishadowing effect in
the structure functions originates dynamically from the gluon
fusions in our model, while in the DGLAP analysis the effect is
partly from the input distributions of the valence quarks [27,28].
Another reason is that the $A$-dependent nonlinear terms enhance the
effect of the gluon fusion in the heavy nuclei.

\newpage
\begin{center}
\section{The Cronin effect}
\end{center}

    The Cronin effect is described by the nuclear modification
factor $R_{dA}$, which is defined as the ratio of the number of
particles produced in a $d+A$ collision to that in a $p+p$ collision scaled by the number of collisions

$$R_{dA}(k_t)=\left.\frac{\frac{dN_{d-A}(k_t,\eta)}
{d^2k_Td\eta}}{N_{coll}\frac{dN_{p-p}(k_t,\eta)}{d^2k_td\eta}}\right\vert_{\eta=0},\eqno(35)$$
$k_t$ and $\eta$ are respectively the transverse momentum and the
pseudo-rapidity of the observed hadron. $N_{coll}$ is
the number of collisions in $d+A$ scattering. In Eq. (35),

$$\frac{dN_{p-p}(k_t,\eta)}{d^2k_td\eta }=\frac{1}{\sigma_{in}}\frac{d \sigma_{p-p}(k_t,\eta)}{d^2k_t d\eta}$$
$$=\frac{1}{\sigma_{in}}\int\frac{dz}{z}J(\eta;k_t;m_{eff})
\left.D_p(z)\delta^2(k_t-zk_{t,g})\frac{d
\sigma_{p-p}(k_{t,g},y)}{dy d^2k_{t,g}}\right|_{y\rightarrow\eta},
\eqno(36)$$and

$$\frac{1}{N_{coll}}\frac{dN_{d-A}}{d^2k_td\eta }=\frac{1}{\sigma_{in}}\frac{d \sigma_{d-A}(k_t,\eta)}{d^2k_t d\eta}$$
$$=\frac{1}{\sigma_{in}}\int\frac{dz}{z}J(\eta;k_t;m_{eff})
\left.D_A(z)\delta^2(k_t-zk_{t,g})\frac{d
\sigma_{d-A}(k_{t,g},y)}{dy
d^2k_{t,g}}\right|_{y\rightarrow\eta},\eqno(37)$$ where
$z=k_t/k_{t,g}$; $D_p(z)$ and $D_A(z)$ are the fragmentation
functions of gluon jets in proton and nucleus, where the
factorized scale-dependence of the fragmentation functions are
neglected; The rapidity $y$ of the produced gluon in the
center-of-mass frame of $p+p$ collisions is defined by

$$x_{1/2}=\frac{k_{t,g}}{\sqrt{s}}\cdot\exp(\pm y);\eqno(38)$$
The relation between the rapidity $y$ and pseudorapidity $\eta$ of
massive particles is

$$y=\frac{1}{2}\ln \left[\frac{\sqrt{\frac{m^2_{eff}+p^2_t}{p^2_t}+\sinh^2 \eta}+\sinh \eta}
{\sqrt{\frac{m^2_{eff}+p^2_t}{p^2_t}+\sinh^2\eta}-\sinh \eta}
\right], \eqno(39)$$ where $m_{eff}$ is the typical invariant mass
of the gluon jet.

    We assume that the hadrons in the central region are produced from the hadronization of the gluons in
    $gg\rightarrow g$ mechanism. According to Ref. [34] we have

$$\left.\frac{d\sigma_{p-p}(k_t,\eta)}{d^2k_td\eta}\right\vert_{\eta=0}$$
$$=\frac{4N_c}{N_c^2-1}\int \frac{dz}{z^2}\frac{\alpha_s}{k_{t,g}^2}J^2 D_p(z)$$
$$\left.\int^{k^2_{t,g}} dq^2_{t,g}f_g^p(x,(k_{t,g}-q_{t,g})^2)
f_g^p(x,q_{t,g}^2)\right\vert_{k_{t,g}=Jk_t/z}$$
$$\simeq\frac{4N_c}{N_c^2-1}\int \frac{dz}{z^2}\frac{\alpha_s}{k_{t,g}^2}J^2
D_p(z)$$
$$\left[f_g^p(x,k^2_{t,g})G^p(x,k_{t,g}^2)+G^p(x,k^2_{t,g})f^p_g(x,k_{t,g}^2)
\right]_{k_{t,g}=Jk_t/z},\eqno(40)$$

$$\left.\frac{d\sigma_{d-A}(k_t,\eta)}{d^2k_td\eta}\right\vert_{\eta=0}$$
$$=\frac{4N_c}{N_c^2-1}\int \frac{dz}{z^2}\frac{\alpha_s}{k_{t,g}^2}J^2 D_A(z)$$
$$\left.\int^{k^2_{t,g}} dq^2_{t,g}f_g^p(x,(k_{t,g}-q_{t,g})^2)
f_g^A(x,q_{t,g}^2)\right\vert_{k_{t,g}=Jk_t/z}$$
$$\simeq\frac{4N_c}{N_c^2-1}\int \frac{dz}{z^2}\frac{\alpha_s}{k_{t,g}^2}J^2
D_A(z)$$
$$\left[f_g^p(x,k^2_{t,g})G^A(x,k_{t,g}^2)+G^p(x,k^2_{t,g})f^A_g(x,k_{t,g}^2)
\right]_{k_{t,g}=Jk_t/z},\eqno(41)$$ here we neglect the
$k_t$-dependence in the fragmentation functions.

     At the first step, we neglect the interactions at final state,
i.e., in Eqs. (40) and (41)

$$D_p(z)=D_A(z)=\delta(z-1). \eqno(42)$$

    We indicate this nuclear modification factor as $R^g_{dA}$, which is drawn in Fig.
9. According to Eq. (38) at $y=0$ and $\sqrt{s}=200GeV$, the
antishadowing effect on the gluon jet should distribute in a broad
range $8 GeV<k_t<60 GeV$, which corresponds to the antishadowing
range $0.02<x<0.3$ in Fig.4.

    At the next step, we consider the corrections of the
fragmentation functions but neglect the difference between proton
and nucleus. We take [66]

$$D_p(z)=D_A(z)=\frac{2}{3}\times1940z^{1.4}(1-z)^8,\eqno(43)$$ where
$D(z)\rightarrow0$ at $z\rightarrow 0$ since the coherence effects
in QCD at small z. Our results are plotted with the solid curve in
Fig. 10. The data are taken from the BRAHMS results in [6]. One can
find that the fragmentation functions shift cross point between the
Cronin and anti-Cronin effects towards small $k_t$, since the
position of the peak value of the fragmentation functions always
localizes at small $z$. We find that the nuclear shadowing and
antishadowing effects at the initial state dominate the Cronin
effect, although a small nuclear modification to the fragmentation
functions, i.e., $D_A(z)\ne D_p(z)$ is necessary.

   The study on the parton energy loss caused by medium-induced
multiple gluon emission in various nuclear conditions is a hot
topic, since the data of $Au+Au$ collisions at RHIC show a new hot
matter which might be a strongly coupled Quark-Gluon Plasma (sQGP).
The presence of a dense medium influences the space-time development
of the partonic shower of a jet. When an energetic jet of parton
propagates through the medium, a part of its energy transfers to the
thermal partons, which is called the parton energy loss. After this
jet propagates a long distance in an expanding de-confinement
system, most of the gluons carrying the lost energy will escape from
the jet cone and will be un-measured. Thus, as an example, the
modified fragmentation function in an effective model can be written
as [67,68]

$$D_A(z)=\frac{1}{1-\epsilon}D_p(\frac{z}{1-\epsilon}),  \eqno(44)$$
where E is the initial energy of a gluon jet and $\epsilon=\Delta E/E$ is fractional energy loss.

    We consider a similar energy loss mechanism which exits in $d+Au$
collisions at RHIC but with a smaller value of $\epsilon$. The
dashed curve plotted in Fig. 10 is the resulting nuclear suppression
factor with $\epsilon=0.1$. From the results motioned above we find
that the anti-Cronin suppression and Cronin enhancement originate
from the nuclear shadowing and antishadowing effects in the initial
state in the EMC effect.

\newpage
\begin{center}
\section{The signals of QGP}
\end{center}

  One of the important findings at RHIC and LHC is that high transverse momentum hadron production
in central heavy ion collisions is suppressed compared with that in
(properly scaled) p+p collisions [1, 2]. This suppression is defined
as the nuclear suppression factor $R_{AA}$ and can be attributed to
energy loss of high-$k_t$ partons that traverse through the hot and
dense medium formed in these collisions. However, to extract the
energy loss from $R_{AA}$, it is necessary to have the nuclear
effects in the initial parton distributions. Such information can be
derived from the integrated and unintegrated gluon distributions in
proton and nuclei we obtained.

    Several of the most important results obtained at RHIC are related to the high-$k_t$ spectrum in
heavy ion $Au+Au$ collisions. In this aspect, a precise
determination of the nuclear effects in the initial state of these
collisions is fundamental. We present the estimation of the ratio
$R_{AA}$ using our parameters in explanation of the EMC- and
Cronin-effects

$$R_{AA}(k_t)=\left.\frac{\frac{dN_{A-A}(k_t,\eta)}
{d^2k_Td\eta}}{N_{coll}\frac{dN_{p-p}(k_t,\eta)}{d^2k_td\eta}}\right\vert_{\eta=0},\eqno(45)$$

Firstly we assume that $\epsilon=0$ in Eq. (44) and calculate the
nuclear shadowing and antishadowing effects in the nuclear
suppression factor. The result is plotted with the dashed curve in
Fig. 11. There exists a big difference between the curve and the
data at $k_t>3GeV$ , which is commonly interpreted in terms of a
strong energy loss of the energetic partons when they traverse
through a dense medium. For example, we take $\epsilon=0.4$ (see the
dotted curve in Fig. 11). Obviously, a true form of fractional
energy loss $\epsilon$ has a crossover from a small value of
$\epsilon$ to a large one when the energy of the gluon jet is
increasing. It is interesting that when we take

$$\epsilon=\left\{ \begin{array}{ll}a &{\rm if }~E< E_c\\
b&{\rm if}~ E> E_c\end{array}\right., \eqno(46)$$ where $a=0.2$ and
$b=0.4$ (i.e., Fig. 12a), we have

$$\left.\frac{d\sigma_{A-A}(k_t,\eta)}{d^2k_td\eta}\right\vert_{\eta=0}$$
$$=\frac{4N_c}{N_c^2-1}\int \frac{dz}{z^2}\frac{\alpha_s}{k_{t,g}^2}J^2 D_A(z)$$
$$\left.\int^{k^2_{t,g}} dq^2_{t,g}f_g^A(x,(k_{t,g}-q_{t,g})^2)
f_g^A(x,q_{t,g}^2)\right\vert_{k_{t,g}=Jk_t/z}$$
$$\simeq\frac{4N_c}{N_c^2-1}\int \frac{dz}{z^2}\frac{\alpha_s}{k_{t,g}^2}J^2
D_A(z)$$
$$\left[f_g^A(x,k^2_{t,g})G^A(x,k_{t,g}^2)+G^A(x,k^2_{t,g})f^A_g(x,k_{t,g}^2)
\right]_{k_{t,g}=Jk_t/z},\eqno(47)$$

$$\left.\frac{d\sigma_{A-A}(k_t,\eta)}{d^2k_td\eta}\right\vert_{\eta=0}$$
$$=\frac{4N_c}{N_c^2-1}\int_0^{k_t/E_c} \frac{dz}{z^2}\frac{\alpha_s}{k_{t,g}^2}J^2
D_A(z)
\left[f_g^A(x,k^2_{t,g})G^A(x,k_{t,g}^2)+G^A(x,k^2_{t,g})f^A_g(x,k_{t,g}^2)
\right]_{k_{t,g}=Jk_t/z}$$
$$+\frac{4N_c}{N_c^2-1}\int_{k_t/E_c}^{1-b} \frac{dz}{z^2}\frac{\alpha_s}{k_{t,g}^2}J^2
D_A(z)
\left[f_g^A(x,k^2_{t,g})G^A(x,k_{t,g}^2)+G^A(x,k^2_{t,g})f^A_g(x,k_{t,g}^2)
\right]_{k_{t,g}=Jk_t/z}$$
$$if ~k_t<E_c(1-b);$$
$$=\frac{4N_c}{N_c^2-1}\int_0^{1-a} \frac{dz}{z^2}\frac{\alpha_s}{k_{t,g}^2}J^2
D_A(z)
\left[f_g^A(x,k^2_{t,g})G^A(x,k_{t,g}^2)+G^A(x,k^2_{t,g})f^A_g(x,k_{t,g}^2)
\right]_{k_{t,g}=Jk_t/z}$$
$$+\frac{4N_c}{N_c^2-1}\int_{k_t/E_c}^{1-b} \frac{dz}{z^2}\frac{\alpha_s}{k_{t,g}^2}J^2
D_A(z)
\left[f_g^A(x,k^2_{t,g})G^A(x,k_{t,g}^2)+G^A(x,k^2_{t,g})f^A_g(x,k_{t,g}^2)
\right]_{k_{t,g}=Jk_t/z}$$
$$if ~E_c(1-b)<k_t<E_c(1-a);$$
$$=\frac{4N_c}{N_c^2-1}\int_0^{1-a} \frac{dz}{z^2}\frac{\alpha_s}{k_{t,g}^2}J^2
D_A(z)
\left[f_g^A(x,k^2_{t,g})G^A(x,k_{t,g}^2)+G^A(x,k^2_{t,g})f^A_g(x,k_{t,g}^2)
\right]_{k_{t,g}=Jk_t/z} $$
$$if~k_t>E_c(1-a).  \eqno(48)$$  we find that the resulting solid curve plotted in Fig.
11 consists with the experiment data at $1 GeV<k_t<10 GeV$.

   Recently, the nuclear modification factor in central $Pb+Pb$ collisions at $\sqrt{s}
= 2.76 TeV$ is published by the ALICE Collaboration at LHC [2]. The
data indicate that $R_{AA}$ reaches a minimum at $k_t=6-7GeV$, which
is about 0.14 and smaller than that at RHIC. However, it rises
steeply the asymptotic value of RHIC at large $k_t$. Therefore, it
is unclear whether a more dense mater is formed or not at LHC.
Obviously, it is necessary to determinate quantitatively the energy
loss after gluon shadowing and antishadowing effects are excluded.
For this sake, similar to the above approach, we take $a=0$ and
$b=0.58$ (Fig. 12b) in Eq. (46) and plot the result with the solid
curve in Fig. 13. The dashed and pointed curves are plotted
respectively when $\epsilon=0$ and $0.58$. Equation (46) is a good
approximation to describe the nuclear suppression factor at
$1GeV<k_t<10 GeV$ although the results show that the energy loss
$\epsilon$ decrease with the jet energy $E>>10~GeV$. Thus, a rapid
crossover from weak energy loss to strong energy loss at a universal
critical energy of gluon jet $E_c\sim 10 GeV$ is found.

    Finally, we discuss the contributions of the BFKL-corrections. When Eq. (29) is neglected, it is found that
the BFKL-corrections to the ratios $R_{dA}$ and $R_{AA}$ can be
neglected in the present energy scale.

     In summary, the EMC- and Cronin-effects are explained by a
unitarized evolution equation, where the shadowing and antishadowing
corrections are dynamically produced by gluon fusions. An alternative form of the GLR-MQ-ZRS equation is derived. The
resulting integrated and unintegrated gluon distributions in proton and nuclei are
used to analyze the contributions of the initial parton
distributions to the nuclear suppression factor in heavy ion
collisions. A simulation of the fractional energy loss
is extracted from the data at RHIC and LHC, where the
contributions of the nuclear shadowing and antishadowing effects are
considered. A rapid crossover from weak energy loss to strong energy loss is found at a universal critical energy of gluon jet $E_c\sim 10 GeV$.

\vspace{0.3cm}

\noindent {\bf Acknowledgments}: This work was supported in part by the
National Natural Science Foundations of China (under Grants No. 10875044 and No. 11205227) and
the Project of Knowledge Innovation Program (PKIP) of Chinese Academy of Sciences, Grant No. KJCX2.YW.W10

\newpage
\noindent {\bf Appendix. GLR-MQ-ZRS equation}: The modifications
of the gluon recombination to the DGLAP evolution in the
GLR-MQ-ZRS equation has following form [38-40], which work in
whole $x$ range .

$$\frac{dG(x,Q^2)}{d\ln Q^2}$$
$$=P^{AP}_{gg}\otimes G(x,Q^2) + P^{AP}_{gq}\otimes S(x,Q^2)  $$
$$-2\int_{x}^{1/2}\frac{dx_1}{x_1}\frac{x}{x_1}
{\cal K}_{GLR-MQ-ZRS}^{GG\rightarrow GG,~
LL(Q^2)}\left(\frac{x}{x_1},\alpha_s\right) G^{(2)}(x_1,\mu_1^2)$$
$$+\int_{x/2}^{1/2}\frac{dx_1}{x_1}\frac{x}{x_1} {\cal
K}_{GLR-MQ-ZRS}^{GG\rightarrow GG,~
LL(Q^2)}\left(\frac{x}{x_1},\alpha_s\right)
G^{(2)}(x_1,\mu_1^2),\eqno(A-1)$$ for gluon distribution and

$$\frac{dxS(x,Q^2)}{d\ln Q^2}$$
$$=P^{AP}_{qg}\otimes G(x,Q^2) + P^{AP}_{qq}\otimes S(x,Q^2) $$
$$-2\int_{x}^{1/2}\frac{dx_1}{x_1}\frac{x}{x_1}
{\cal K}_{GLR-MQ-ZRS}^{GG\rightarrow S\overline{S},~
LL(Q^2),}\left(\frac{x}{x_1},\alpha_s\right) G^{(2)}(x_1,\mu_1^2)$$
$$+\int_{x/2}^{1/2}\frac{dx_1}{x_1}\frac{x}{x_1} {\cal
K}_{GLR-MQ-ZRS}^{GG\rightarrow S\overline{S},~
LL(Q^2)}\left(\frac{x}{x_1},\alpha_s\right) G^{(2)}(x_1,\mu_1^2),
\eqno(A-2)$$ for sea quark distributions, where $P^{AP}$ are the
evolution kernels of the linear DGLAP equation and the recombination
functions

$$\frac{dx_1}{x_1}{\cal K}_{GLR-MQ-ZRS}^{GG\rightarrow GG,~LL(Q^2)}$$
$$=\frac{9\alpha^2_s}{64}\frac{(2x_1-x)(72x_1^4-48x_1^3x+140x_1^2x^2-116x_1x^3+29x^4)}{x_1^5x}dx_1
\eqno(A-3)$$

$$\frac{dx_1}{x_1}{\cal K}_{GLR-MQ-ZRS}^{GG\rightarrow S\overline{S},~LL(Q^2)}$$
$$=\frac{\alpha^2_s}{48}\frac{(2x_1-x)^2(18x^2_1-21x_1x+14x)}{x_1^5}dx_1
\eqno(A-4)$$

\newpage
\vskip 3.0 truecm \hbox{
\centerline{\epsfig{file=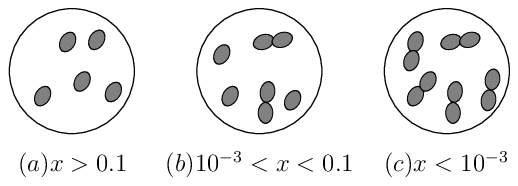,width=15.0cm,clip=}}}

\vskip -12.0 truecm
 Fig. 1 The kinematic regions of the DGLAP- and BFKL
equations. \vskip 1.0 truecm

\newpage
\vskip 1.0 truecm \hbox{
\centerline{\epsfig{file=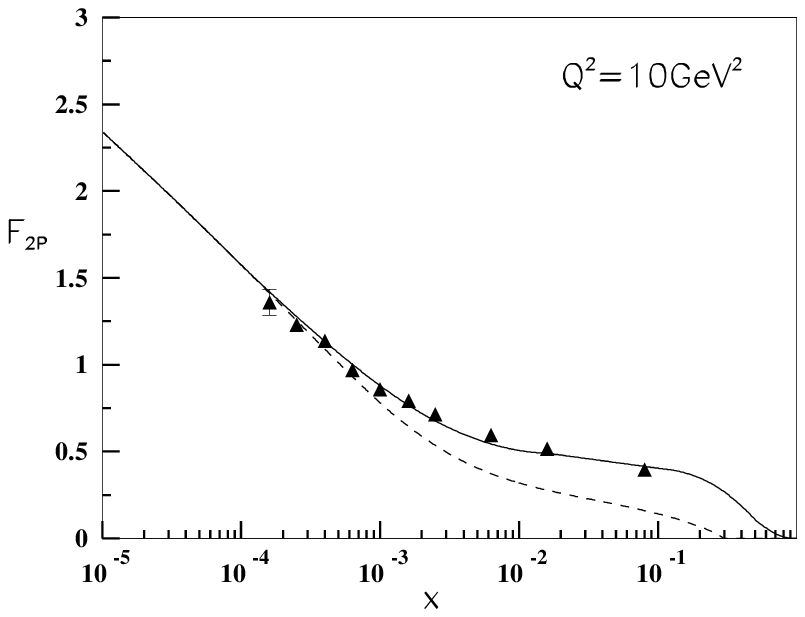,width=15.0cm,clip=}}} \vskip
-4.0 truecm
 Fig. 2  The fit of the computed $F_{2P}(x,Q^2=10~GeV^2)$ in proton
by the evolution equations (21), (29) and (32) using the input Eq.
(33) (dashed curve). The contributions of the valence quarks are
parameterized by the differences between solid and dashed curves.
The data are taken from Ref. [58].\vskip 1.0 truecm

\newpage
\vskip 1.0 truecm \hbox{
\centerline{\epsfig{file=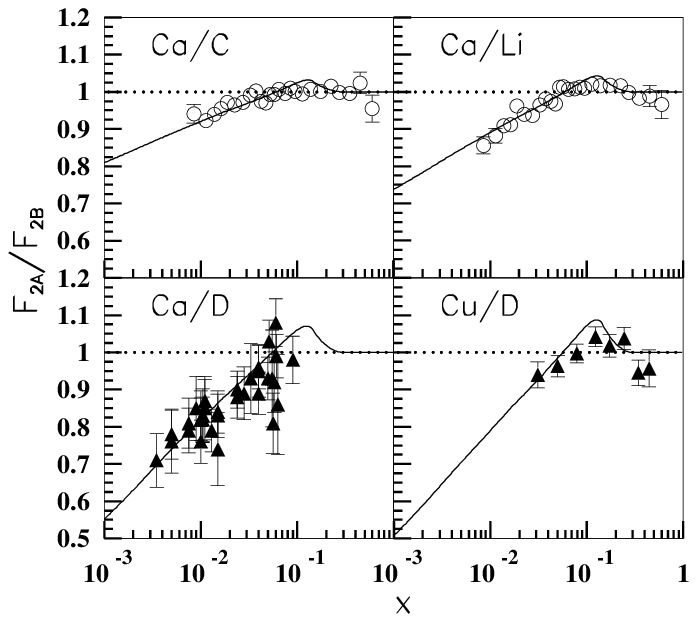,width=15.0cm,clip=}}} \vskip
-4.0 truecm Fig. 3 Predictions of the evolution equations (21),
(22) and (29) compared with the EMC ratio of the structure
functions for various nuclei. The data are taken from [59-62]. All
curves are for $Q^2=10~GeV^2$. \vskip 1.0 truecm

\newpage
\vskip 1.0 truecm \hbox{
\centerline{\epsfig{file=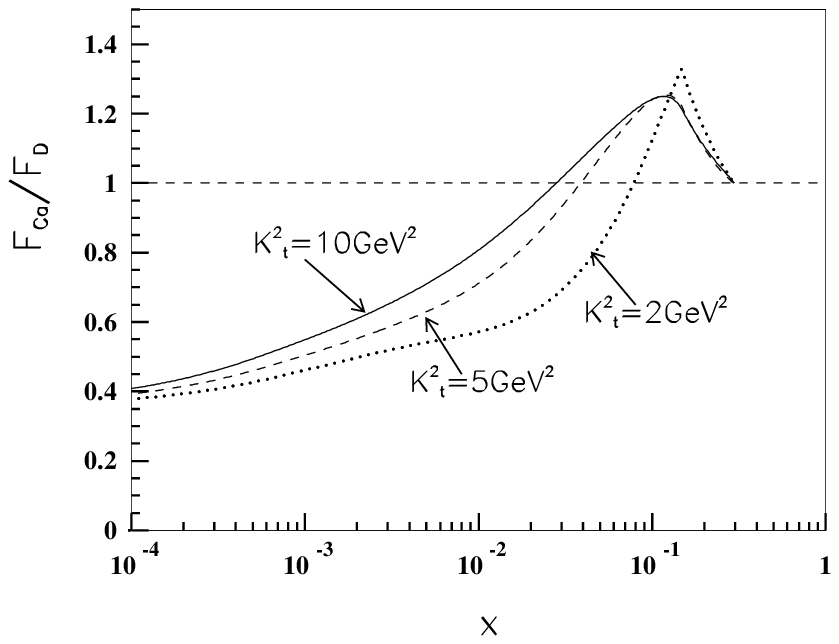,width=15.0cm,clip=}}} \vskip
-4.0 truecm Fig. 4 Predictions of the evolution equations (21),
(22) and (29) for the ratio of the unintegrated gluon
distributions in Ca/D with different values of $x$ and given
$k_t$. \vskip 1.0 truecm

\newpage
\vskip 1.0 truecm \hbox{
\centerline{\epsfig{file=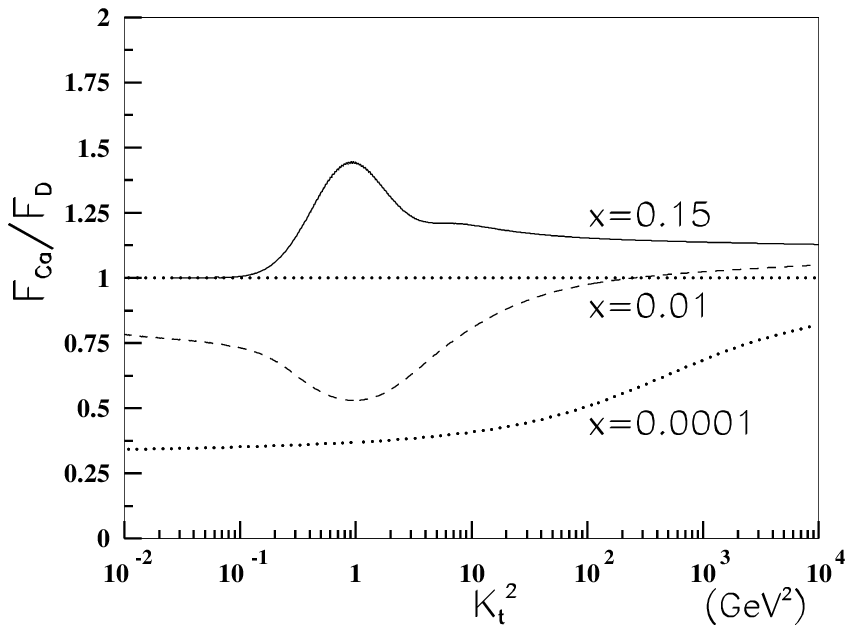,width=15.0cm,clip=}}} \vskip
-4.0 truecm
 Fig. 5
Similar to Fig. 4 but with different values of $k_t$ and given
$x$. \vskip 1.0 truecm

\newpage
\vskip 1.0 truecm \hbox{
\centerline{\epsfig{file=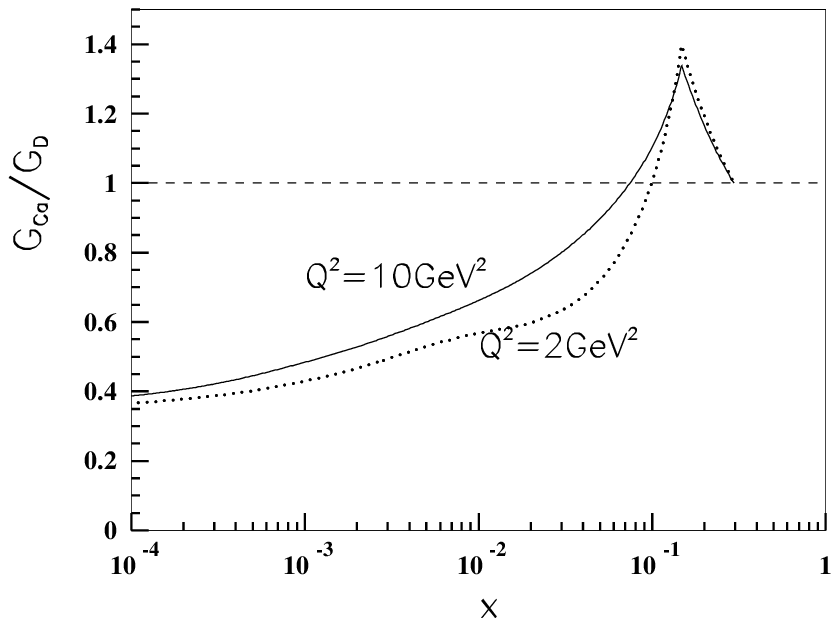,width=15.0cm,clip=}}} \vskip
-4.0 truecm Fig. 6 $x$-dependence of the ratio for the integrated
gluon distributions in $Ca/D$ with the evolution equations (21),
(22) and (29). \vskip 1.0 truecm

\newpage
\vskip 1.0 truecm \hbox{
\centerline{\epsfig{file=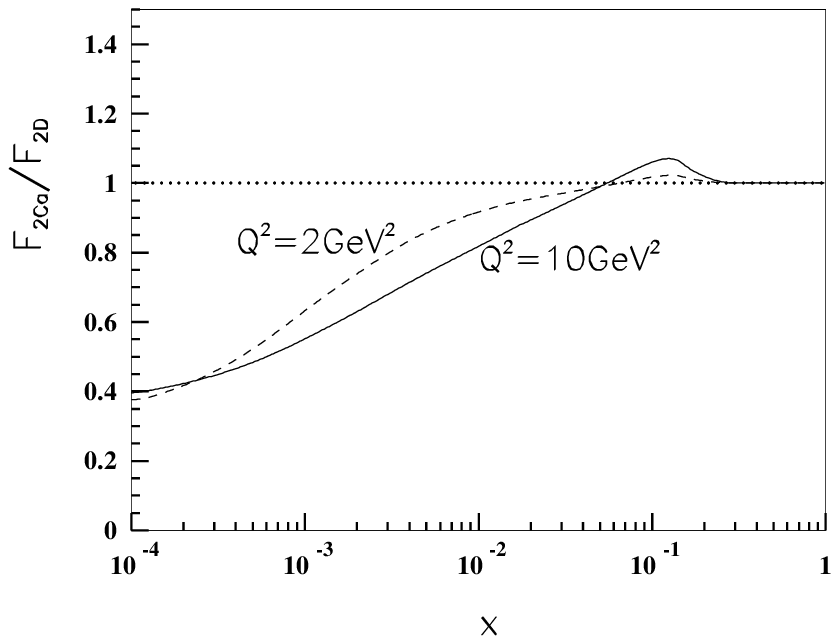,width=15.0cm,clip=}}} \vskip
-4.0 truecm Fig. 7 Similar to Fig. 6 but for the the ratio of the
structure functions. \vskip 1.0 truecm

\newpage
\vskip 1.0 truecm \hbox{
\centerline{\epsfig{file=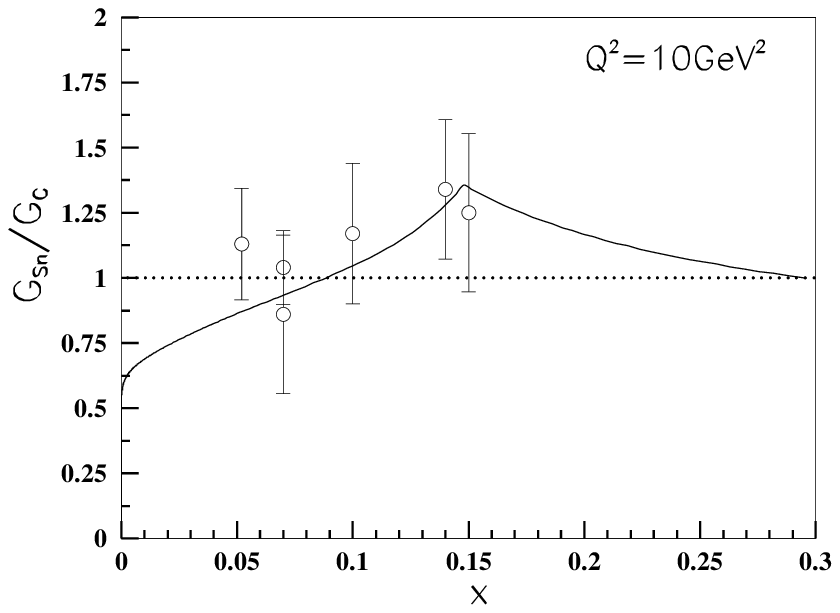,width=15.0cm,clip=}}}

\vskip -4.0 truecm Fig. 8 Predictions for the ratio of the gluon
distributions in $Sn/C$ and the data are taken from Ref. [64].
\vskip 1.0 truecm

\newpage
\vskip 1.0 truecm \hbox{
\centerline{\epsfig{file=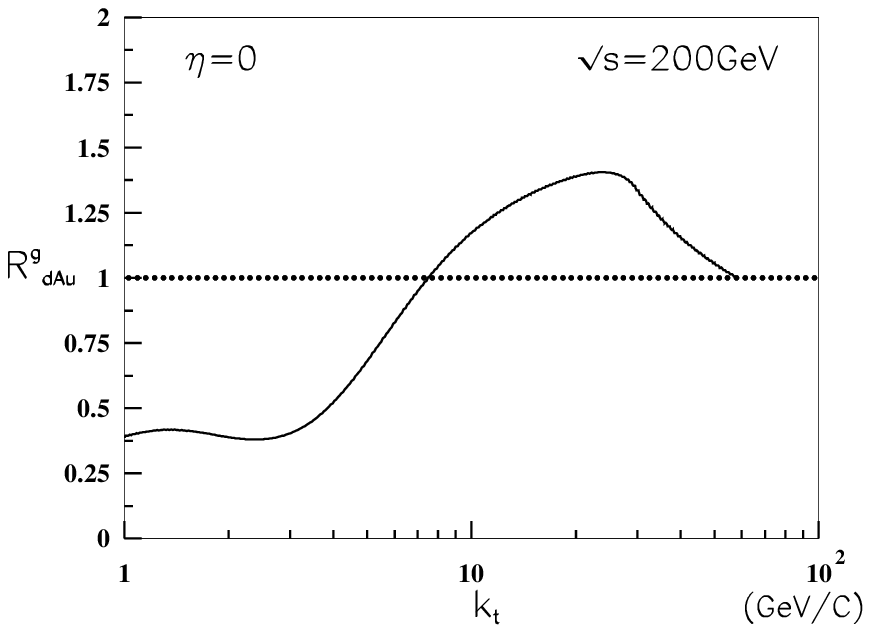,width=15.0cm,clip=}}} \vskip
-4.0 truecm Fig. 9 Predicted nuclear modification factor
$R^g_{dAu}$ of gluon jet in central $d+Au$ collisions at $\sqrt
{s}=200 GeV$. \vskip 1.0 truecm

\newpage
\vskip 1.0 truecm \hbox{
\centerline{\epsfig{file=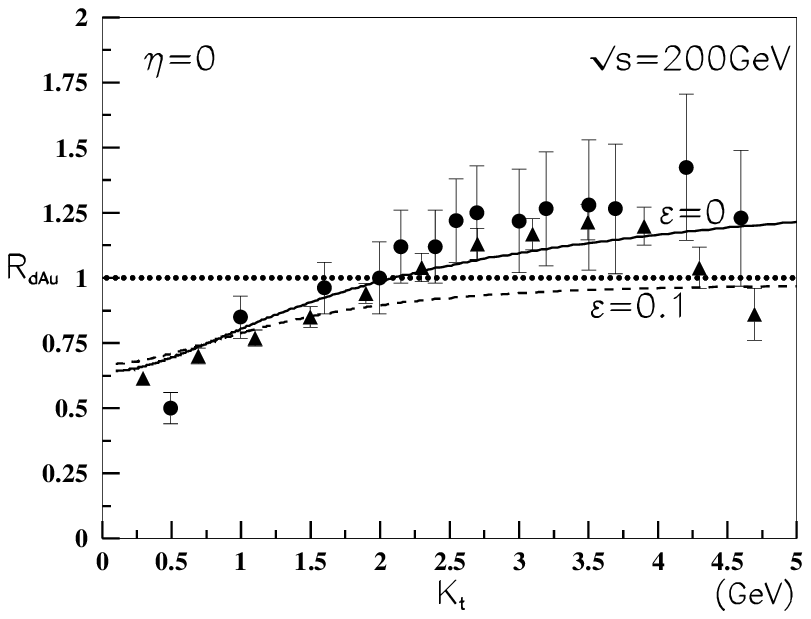,width=15.0cm,clip=}}} \vskip
-4.0 truecm
 Fig. 10
Nuclear modification factor $R_{dAu}$ of charged particles in
central $d+Au$ collisions at $\sqrt {s}=200 GeV$, where fractional
energy loss $\epsilon=0$ (solid curve) and $0.1$ (dashed curve).
The data are taken from Ref. [1]. \vskip 1.0 truecm

\newpage
\vskip 1.0 truecm \hbox{
\centerline{\epsfig{file=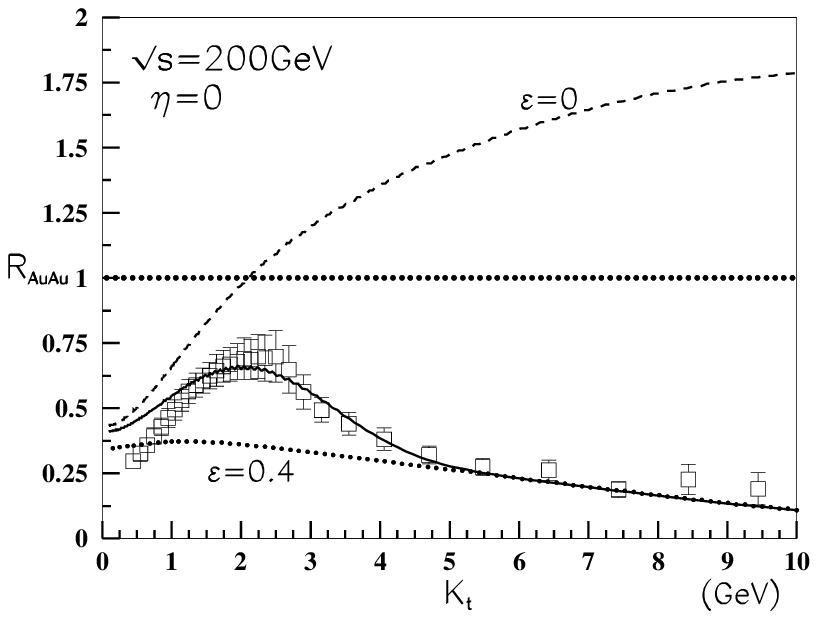,width=15.0cm,clip=}}} \vskip
-4.0 truecm Fig. 11 Estimated nuclear suppression factor $R_{AA}$
in central $Au+Au$ collisions at $\sqrt{s}=200GeV$: solid curve
using Eq. 46 with $a=0.2$, $b=0.4$, (see Fig. 12a); dashed curve
using $\epsilon=0$, and pointed curve using $\epsilon=0.4$. The
data are taken from Ref. [1]. \vskip 1.0 truecm

\newpage
\vskip 1.0 truecm \hbox{
\centerline{\epsfig{file=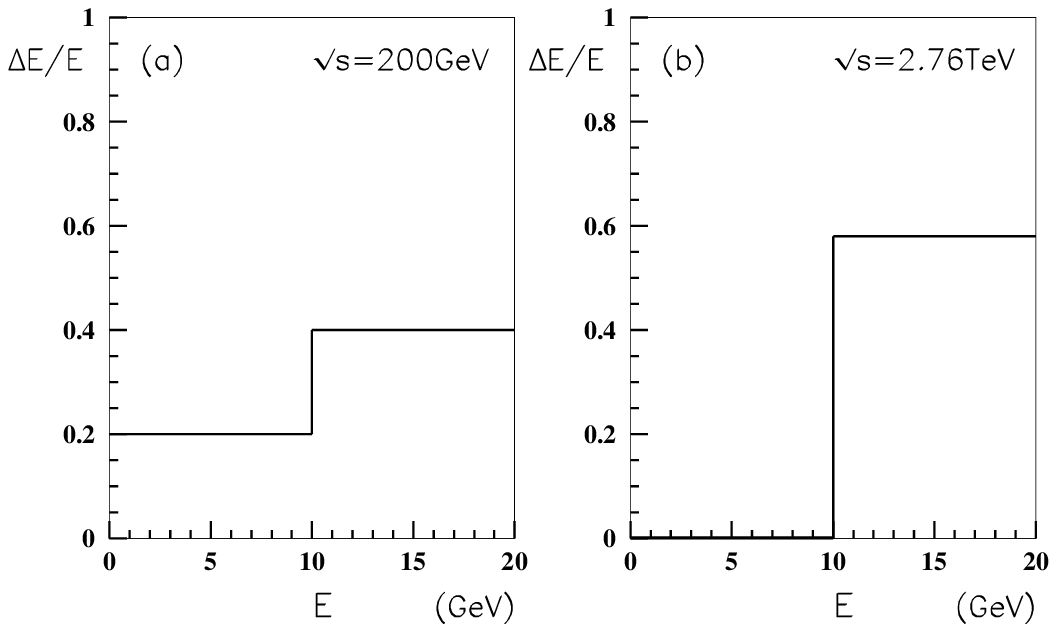,width=15.0cm,clip=}}} \vskip
-4.0 truecm Fig. 12 Two possible fractional energy losses, which
correspond to (a) central collisions at $\sqrt{s}=200GeV$ for
$Au+Au$ and (b) at $\sqrt{s}=2.76TeV$ for $Pb+Pb$, respectively.
\vskip 1.0 truecm

\newpage
\vskip 1.0 truecm \hbox{
\centerline{\epsfig{file=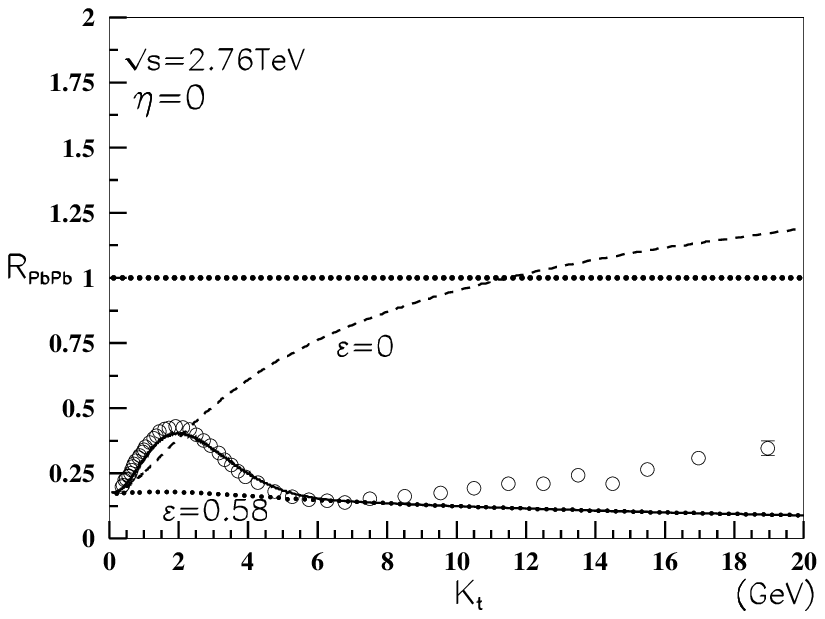,width=15.0cm,clip=}}} \vskip
-4.0 truecm Fig. 13 Similar to Fig. 11, but for central $Pb+Pb$
collisions at $\sqrt{s}=2.76TeV$, where solid curve using Eq. 46
with $a=0$, $b=0.58$, (see Fig. 12b); dashed curve using
$\epsilon=0$, and pointed curve using $\epsilon=0.58$. The data
are taken from Ref. [2]. \vskip 1.0 truecm

\end{document}